# PKI IN GOVERNMENT IDENTITY MANAGEMENT SYSTEMS

Ali M. Al-Khouri

Emirates Identity Authority, Abu Dhabi, United Arab Emirates.
ali.alkhouri@emiratesid.ae

## ABSTRACT

*The purpose of this article is to provide an overview of the PKI project initiated part of the UAE national ID card program. It primarily shows the operational model of the PKI implementation that is indented to integrate the federal government identity management infrastructure with e-government initiatives owners in the country. It also explicates the agreed structure of the major components in relation to key stakeholders; represented by federal and local e-government authorities, financial institutions, and other organizations in both public and private sectors. The content of this article is believed to clarify some of the misconceptions about PKI implementation in national ID schemes, and explain how the project is envisaged to encourage the diffusion of e-government services in the United Arab Emirates. The study concludes that governments in the Middle East region have the trust in PKI technology to support their e-government services and expanding outreach and population trust, if of course accompanied by comprehensive digital laws and policies.*

## 1. INTRODUCTION

Many countries around the world have invested momentously in the development and implementation of e-government initiatives in the last decade. As of today, more and more countries are showing strong preference to develop "the 24-hour authority" [1] and the delivery of further self-service models via digital networks [2]. However, from a citizens angle, and although individuals with higher levels of education are in general more open to using online interactions, there is a stronger preference among the majority for traditional access channels like in person or telephone-based interactions with government and private organisations, [2,3].

Our research shows that governments in most parts of the world have been challenged in gaining citizen engagement in the G2C transactions. Our earlier study pointed to the fact that e-government initiatives around the world have not succeeded to go to the third and forth phases of e-government development [4,5] (see also Figure 1). In this earlier study, we referred to the need of fundamental infrastructure development in order to gaining the trust of citizens, and hence expanding outreach and accelerating e-government diffusion. One of the key components we highlighted there was the development and integration of a government identity management system with PKI technology to enable stronger authentication of online users.

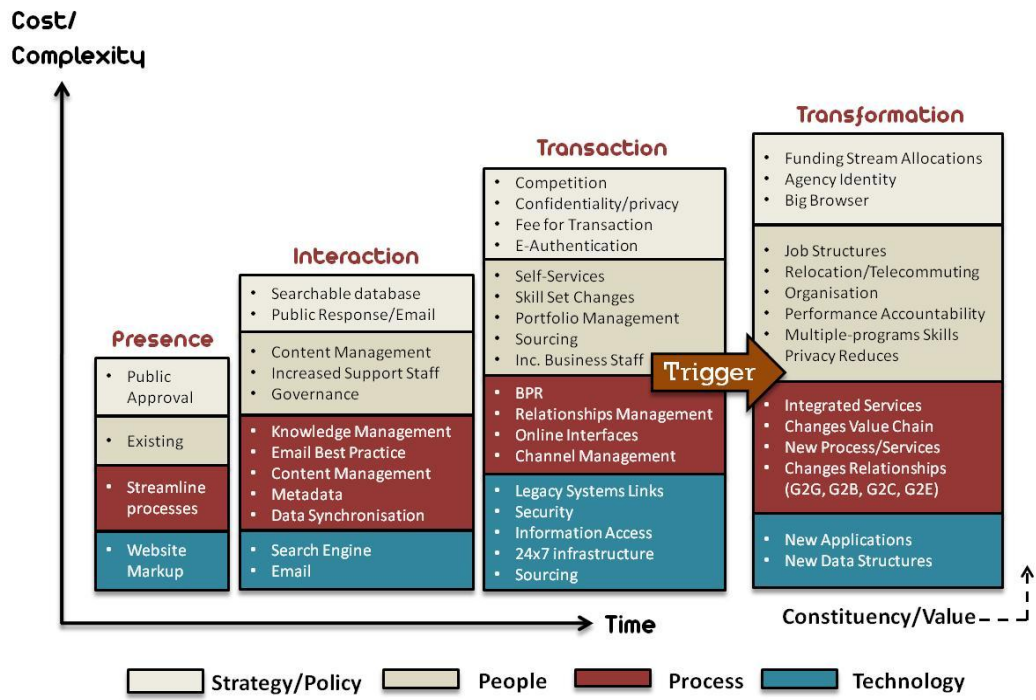

Figure 1. Four Phases of e-Government – [6]

The purpose of this article is to describe the UAE government approach of incorporating PKI into their ID card architecture. It explains the major components of the project related to e-government G2C progress. In doing so, we seek to make a contribution to the available research literature on the implementation of PKI in national identity management systems, and its role in the diffusion of e-government and outreach. This article is structured as follows. Some background introduction on PKI is provided first. The UAE PKI project is introduced next, and a highlight is provided on its major components. Some reflection is provided on key management considerations, before the paper is concluded.

## 2. DIGITAL IDENTITIES AND PKI IN E-GOVERNMENT

For the past ten years, governments around the world have been vitally concerned with the establishment of secure forms of identification and improved identity management systems, in order to ascertain the true identities and legitimacy of their population. Yet, many organizations both in public and private sectors still rely heavily on their own constructed models of relevant online identities, which are based on captured data from single or multiple sources and transforming them into own data structures within their information systems.

With the revolution of digital networks, governments are realizing their roles to develop foundational infrastructure for digital identities. The term "digital identity" refers to a set of attributes and properties about an individual that are associated together and available in an electronic form to construct trusted digital credentials.

Evidently, governments have long played an authoritative role in identity provision in the physical world, and are now faced with demands to establish digital societies and identities in order to support e-government and e-commerce initiatives. It is the role of the government to associate digital identities to specific persons who will be authorised to perform certain actions in physical or digital forms.

This association is facilitated through digital certificates and digital signatures that altogether construct the digital identity [7]. Thus many governments have considered PKI technology to establish and implement this binding through registration and digital certificate issuance process. In basic terms, PKI attaches identities to digital certificates for the purpose of assured, verifiable, and secure digital communications.

Public key infrastructure commonly referred to as PKI is an Information Technology (IT) infrastructure and is a term used to describe the laws, policies, procedures, standards, and software that regulate and control secure operations of information exchange based on public and private keys cryptography [8]. Table 1 summarizes the primary elements that make up the PKI components. The term PKI is used in this article to refer to the comprehensive set of measures needed to enable the verification and authentication of the validity of each party involved in an electronic transaction.

Table 1. Basic PKI Components

| Component | Description |
|---|---|
| Digital Certificates | Electronic credentials, consisting of public keys, which are used to sign and encrypt data. Digital certificates provide the foundation of a PKI. |
| Certification Authoritie(S) – CAs | Trusted entities or services that issue digital certificates. When multiple CAs are used, they are typically arranged in a carefully prescribed order and perform specialised tasks, such as issuing certificates to subordinate CAs or issuing certificates to users. |
| Certificate Policy and Practice Statements | Documents that outline how the CA and its certificates are to be used, the degree of trust that can be placed in there certificates, legal liabilities if the trust is broken, and so on. |
| Certificate Repositories | A directory of services or other location where certificates are stored and published. |
| Certificate Revocation Lists (CRL) | List of certificates that have been revoked before reaching the scheduled expiration date. |

PKI offers high levels of authentication of online users, encryption and digital signatures , which also support the maintenance of elevated echelons of data privacy, streamline workflow and enable access. The cornerstone of the PKI is the concept of private keys to encrypt or digitally sign information. One of the most significant contributions a PKI has to offer is non-repudiation. Non-repudiation guarantees that the parties involved in a transaction or communication cannot later on deny their participation.

PKI, in general, has grown both more slowly and in somewhat different ways than were anticipated [7]. It has had some success stories in government implementations; the largest PKI implementation to date is the Defense Information Systems Agency (DISA) PKI infrastructure for the Common Access Cards program [9]. See Also Appendix-1. Many researchers pointed out the complexity of PKI, and that it is only sound in theoretical terms [10]. We definitely do not agree with those who claim that PKI cannot be practiced and yield effective results. As with any technology, PKI is not without its own security risks due to its complex architectures. Indeed, there is no bullet-proof technology that could provide us with a fault free solution and meet all of our security needs.

In fact, studies conducted by academics and practitioners remain passionate about the promises of PKI to revolutionise electronic transactions (see for example: [4,5,8,11]. Undoubtedly, published studies in the existing literature contributed significantly to the development of the technology and explaining its benefits. Nonetheless, those studies are believed to remain very much handy to technical researchers.

This is to say that although an awful lot of articles were written on this topic, they seem to be written to improve and develop theoretical frameworks while others tackle narrowed technological issues. See for example: [12-22].

Looking at these studies, we note that although such research efforts have been comprehensive in specific areas, they do not assume a standard or a uniformed PKI approach. Interestingly, some researchers pointed to the fact that this field lacks fundamental theories to guide the development of clear path for PKI practice in our world [23].

Others explain lack of adoption and wide failures in PKI industry to be due to not having enough PKI applications with clear business cases to support the roll out of the infrastructure [24]. Therefore, many implementations reported to have produced unnecessary costs when implemented without clear business cases [25]. Apparently, with the increasing complexity, the implementation of PKI systems becomes extremely challenging in light of the limited documented experiences that have included inefficient and short living implementations, with no clear ROI cases.

## 3. MOTIVATION AND EXPECTED CONTRIBUTION OF THE WORK

In the age of supercomputing, secure communication is becoming a need and a necessity. PKI is being recognized as an important security component in digital infrastructures to support authentication, integrity, confidentiality, and non-repudiation. Organisations that deployed PKI have reported substantial economic savings [26]. Although PKI is reported to gaining wide popularity, it is still implemented in its very basic form and protocols e.g., secure sockets layers are the most common application of PKI. From our perspective, existing literature still does not explain important implementation areas for practitioners in the field.

This article was developed to provide insights of PKI implementation in a government context and from a practitioner viewpoint. Our main motivation is to explain how PKI could be diffused in modern national identity management systems to outline and address appropriate security requirements. Our contributions are related to a PKI implementation model from one of the most pioneering governments in technology adoption in the Middle East.

We believe that PKI deployment can evolve double the speed if adopted and owned by governments as trusted third parties. See also [27]. The role of the trusted third parties would be to verify the identities of the parties wishing to engage in a secure online communication. PKI, particularly in combination with smart ID cards, can provide robust user authentication and strong digital signatures.

Existing PKI deployments have limited customers. The use of PKI in ID card schemes for example, would enjoy larger customer base. It is our belief that such systems have the potential to raise the awareness of both governments and citizens trust levels in electronic transactions with such advanced technologies. These technologies are thought to pave the way for government transformation from service delivery perspectives and introduce new communication and service delivery channels that should replace government traditional physical counter interactions. Successful PKI implementation cases would put higher pressures on both government officials and private sector to develop killer applications to revolutinsie public service sectors. In brief, this article does not intend to explain detailed implementation questions, although it can serve as a primer for government officials and researchers who are interested in PKI implementations in government sectors.

## 4. THE UAE PKI PROJECT

Emirates Identity Authority [28] is implementing PKI and a Federated Identity Management (FIM) solution to complement the existing identity management infrastructure and provide extended services to federal and local e-government authorities in the UAE. The project aims to develop a comprehensive and intergraded security infrastructure to enable a primary service of confirmed digital identities of UAE ID card holders on digital networks; primarily on the internet.

The project has two strategic objectives: (1) to enable verification of the cardholder's digital identity; (authentication services) by verifying PIN Code, biometric, and signature certificate and (2) provide credibility (validation services) through the development of a Central Certification Authority. See also Figure 2 below.

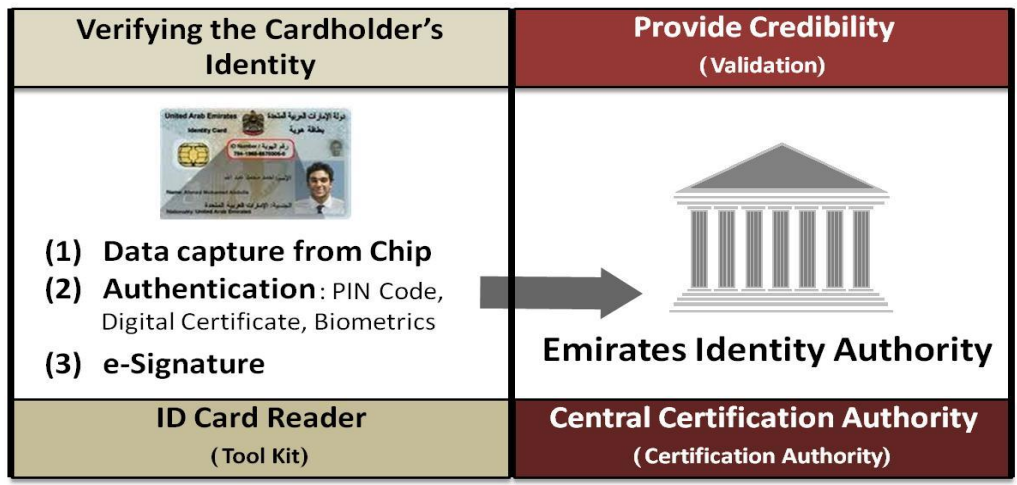

Figure 2. UAE PKI project primary objectives

The PKI project will support the issuance of identity, digital signature and encryption certificates as well as key recovery for private keys associated with encryption certificates. It will also issue different types of certificates to support the requirements of other business sectors and communities. An example of such custom certificates is attribute (role-based) certificates used for example by e-government, healthcare and justice sectors who may require their own role-based access control and administration such as the management of CA permissions, performing specific CA tasks, etc.

Apart from issuing and managing digital certificates, the PKI project will enable business applications to use certificates by making available the proper means to validate PKI-based transactions. It will provide high levels of security infrastructure having the service integrity and assurances required to support the distribution and verification of public key certificates.

We attempt to outline the three main components of the UAE Identity Management Infrastructure project related to G2C e-government:

(1) issuance of smart cards as means to users authentication capability;

(2) card readers and toolkit dissemination to enable smart card applications; and

(3) the development of central certification authority to provide online validation services.

These are discussed next.

## 4.1. Smart Cards and Online Users Authentication

Authentication is the process by which an entity identifies itself prior to network logon is permitted. Smart card authentication is one of the strongest user authentication mechanisms available today in the market. Unlike ordinary cards like those used in banks, smart cards can defend themselves against unauthorized users as it uses complex and high level security measures. Smart cards are considered to represent a breakthrough solution for maximizing security, efficiency and interoperability in a wide range of e-government and e-commerce applications such as strong authentication, identity management, data management, customer support, and communications [29-33].

It is envisaged that the new smart ID card issued by the government to all citizen and resident population in the UAE and as part of the ID card scheme launched in mid-2005 will gradually be the only acceptable token to access e-government portals. Out of an estimated 8.2 million UAE population, more than 3.7 million people already posses smart ID cards, and it is planned that towards the end of 2014 all population will be enrolled.

The latest generation of UAE smart ID cards (144K Contactless) contain multiple credentials, including unique RFID, MRZ barcode, photo ID, and biometric information (fingerprints), along with microprocessor and crypto keys and certificates. There are key primary services that can be provided by the UAE ID card in terms of e-government applications, some features of which are currently in use as explained next (See also Table 2).

Table 2. UAE ID Card Capabilities and Features Basic.

|  | What's in the chip? | How does it work? | Validation? | Applications |
|---|---|---|---|---|
| Identity Data Capture | • Biographical<br>• Photo | • 100% accurate information<br>• Publicly available for applications | • Identity data contained in files **digitally signed by EIDA** | • Provisioning (e.g. Data capture at Bank branches, portal registration) |
| Authentication | • PKI Applet with **authentication key pair** and corresponding certificate | • PIN code protection | • Emirates ID Online/ Offline Validation Services | • Online access to electronic services using the eID as **3 factor authentication** token |
| Signature | • PKI Applet with **signature key pair** and corresponding certificate | • PIN code<br>• eID Digital Signature | • Signature Validation Services | • Specific transactions having **non-repudiation/ legal enforcement** |

- **Trusted Personal Data:** available in an electronic form which allows applications to capture data directly from the smart ID card chip.

    The integration of the capability of reading data electronically from the chip in some public sector applications have introduced significant contributions in terms of speed and accuracy and the elimination of the traditional ways of data capture and entry procedures. The use of the smart ID card for physical authentication and data capture has shortened for example the process cycle of service delivery at one public sector organizations (i.e., Dubai Courts) to less than 7 seconds from 7 to 10 minutes taken previously [34]. Similar success stories were repeated in many of the public sector organisations in the country that contributed to raising awareness of the smart card capabilities.

- **Multi-Factor Authentication:** support varying strengths of authentication i.e., pin code, biometrics, digital certificates.

    The multi-factor authentication feature is a major capability that the ID card provides for e-government applications. For example, Abu Dhabi e-government portal [35] uses the UAE smart ID card to provide higher levels of assurance and confidence in the digital identities that interact with the portal. A two factor authentication (PIN and Offline Certificate validation) capability of the ID card has been integrated to support and enhance the security for different e-service access models.

- **Digital Signature:** personal digital certificates that allow users to digitally sign documents and applications.

  Data integrity and non-repudiation capability of signed documents and applications is another benefit. Since access to the private key component needed to perform digital signatures is restricted to the person who possesses an ID card who has knowledge of its associated user PIN (and biometrics), it becomes increasingly difficult for an individual to later deny (repudiate) participation in transactions involving his or her digital signature. The digital signature is projected to replace the pen-and-ink signatures in both government and private sector transactions. This capability inserted into the card should also further support the development of e-government and e-commerce environments.

## 4.2. Smart Card Development Toolkit / Reader

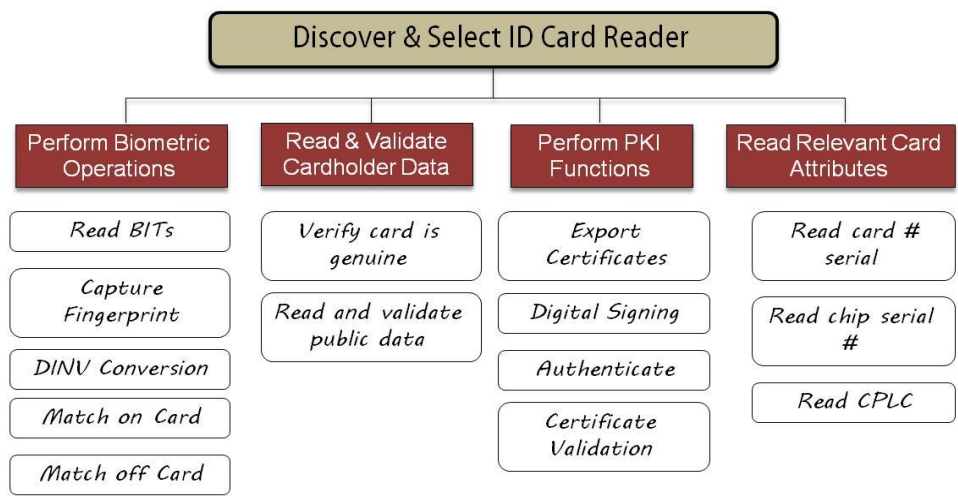

Figure 3. ID Card Toolkit Functions

To use a smart card in an e-government G2C environment, computers are needed to be equipped with smart card readers in order to enable the capabilities specified in section 3.1 above. See also Figure 3. A software toolkit was developed to enable integration with e-government applications which included smart card interface standard and the driver software used for managing the smart card and the card reader. In short, the smart card development toolkit aims to demystify the application of the smart card in e-government transactions, and also strengthen the understanding of all those involved in the planning and the execution of e-government initiatives.

From our research in the field, we noted that smart card manufacturers normally provide their own read and access communication protocols which may raise up some integration limitation issues. Therefore, the development toolkit in the UAE was designed to be free of any proprietary features, and to allow a simple plug-and-play functionality from both the user and service provider perspectives. The type of the reader terminal is dependent on the security access models specified by the service providers. For instance, less sophisticated and cheaper card terminals are available if no biometric authentication is required.

Overall, the developed toolkit was designed to support desktop, client-server, web applications and multiple development environments such as Java, C#, .Net. Figure 4 depicts the internal toolkit structure. See also Appendix-B for further information on the toolkit capabilities supported in offline and online modes.

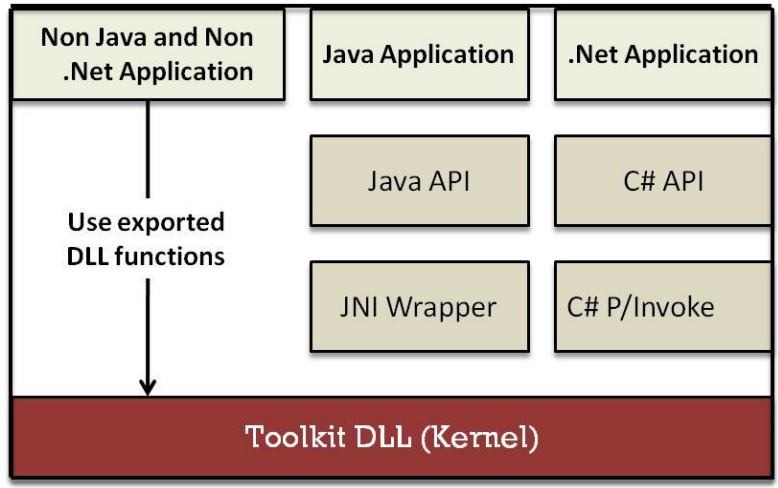

Figure 4. Toolkit Structure

Having said this, the next section explores the major component of this article related to the implementation of a central certification authority in the country and its overall architecture that will be integrated with the above two components to enable online (stronger) credential authentication.

### 3.4. Central Certification Authority

The Central Certification Authority also referred to as the Government Root Certification Authority is intended to be the highest Certification Authority in the hierarchical structure of the Government Public Key Infrastructure in the UAE. The high level UAE PKI architecture depicted in Figure 5 will encompass a root and multiple certified subordinate CAs' to support own PKI policy and function.

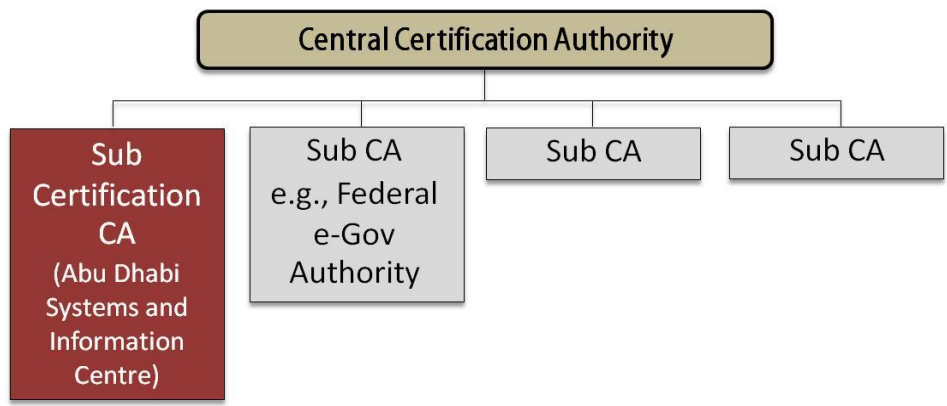

Figure 5. Certifications Authorities Structures

The architectural design of the certification authority infrastructure was discussed and refined at different business and technical levels with key stakeholders representing public and private sectors. The PKI management model was designed to complement existing security management practices followed by those involved in e-government and e-commerce initiatives by providing them with online validation services. As some e-government authorities required their own Certification Authority (CA), it was important for the implemented system to support such requirements. Figure 6 depicts the overall structure of the root CA.

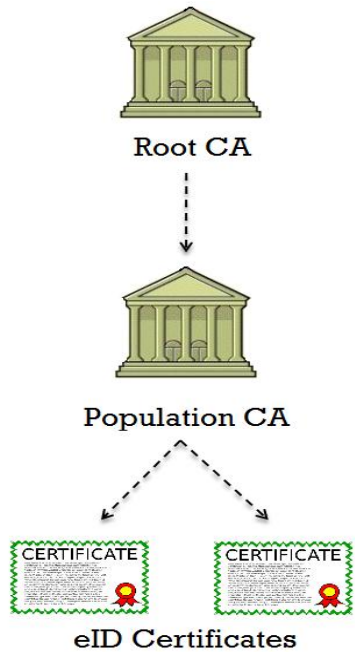

Figure 6. Root CA Structure

The PKI architecture was designed to support two operational models for the implementation of a third party sub CA. In the first option, an e-government authority may implement its own CA including the required software and hardware infrastructure. It will rely on the same PKI infrastructure to certify its Public CA using own Root certificate. Figure 7 below illustrates this solution.

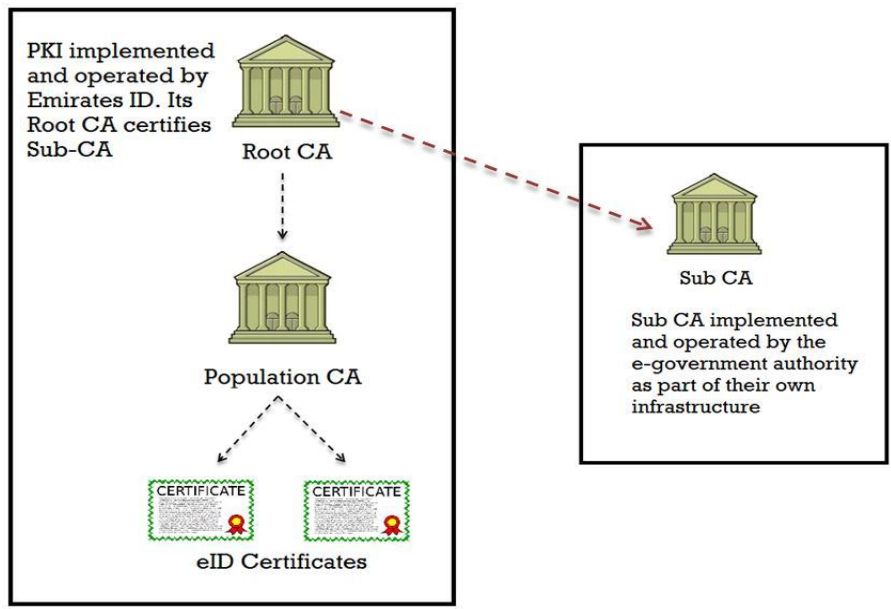

Figure 7. PKI Implementation option 1

The second option assumes that a given e-government authority CA is setup as part of the same PKI infrastructure. A virtual partition is implemented on the Population CA. The e-government CA will be initialized and configured on this new virtual partition. A virtual key container is created on the HSMs so that the Sub CA key pair and corresponding certificates are separated completely from the Root keys. The solution of this second option is illustrated in Figure 8 below.

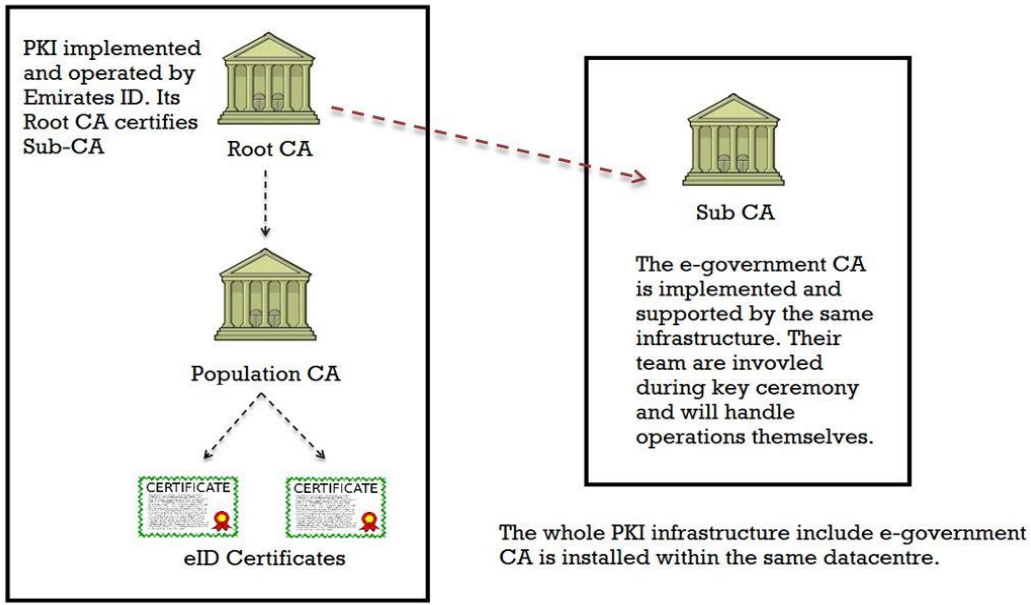

Figure 8. PKI Implementation option 2

Deciding on which option to opt for depends entirely on the e-government authorities' requirements and their readiness to use or operate a PKI infrastructure. The first option meant no particular investments as e-government authorities would rely on the developed PKI infrastructure in the ID Card project to certify their CA public key with their Root CA. The second option involved the implementation and operation of Sub CA by the same root CA authority. The Certificate Policy (CP) needed to be specified for the e-government CA and simply for any CAs certified by the Root CA. The CP which was specified by the Policy Authority described the requirements for the operation of the PKI and granting of PKI credentials as well as the lifetime management of those credentials.

So in practice, after the completion of the authentication process which may include pin and biometrics verification, the transaction is checked for validity. At this stage, and depending on the available infrastructure, a local CRL and/or Certificate Repository database may be consulted. Another cross validation process could take place through connecting to the central certification authority to provide services of authentication and validation. A PKI based workflow depicted in Figure 9 explains how users carrying smart identity cards will interact within a PKI environment.

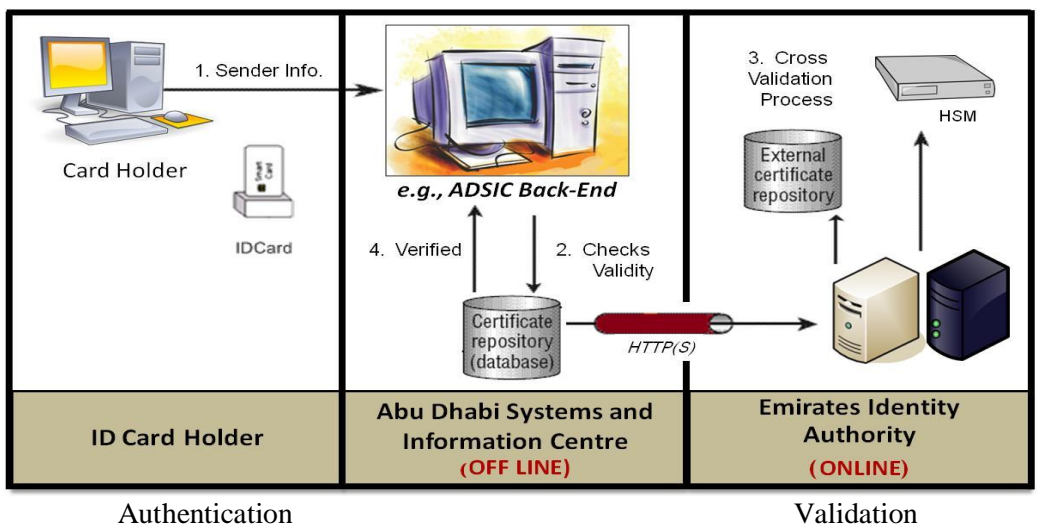

Figure 9. Authentication and Validation (PKI) Workflow

Having said this, the next section attempts to provide a short reflection on some key management considerations to provide guidance to government agencies contemplating the development and deployment of smart ID cards and PKI solutions.

## 5. REFLECTION

### 5.1. Issues related to scalability, operational costs, and integration

Our research on PKI included the evaluation of various commercial software products available in the market. After rigorous benchmarks, the major components of the PKI solution were selected from leading international products. The following three issues needed careful attention:

> **Scalability** - The PKI functionality should scale well to handle millions of certificates and accommodate separate large-scale projects (such as the upcoming UAE biometric e-Passport project, e-Gate project at airports, e-Services project by the various e-government authorities in the country, etc.).
>
> **Operational Costs** - Certificate Authorities and Repositories will need continual operations and maintenance, especially with the increased number of customers and large-scale projects to be supported. PKI structure options should pinpoint associated costs of operations and maintenance.
>
> **Integration** – Applications using PKI (i.e. PKI-enabled applications) shall integrate with central certification authority systems which shall answer the following questions:
>
> - What is the best integration model we can offer to PKI-enabled applications?
> - How can such integration with external applications be performed so that a high degree of security can be guaranteed against unauthorized access?

Stavrou [36] identified five key risks associated with PKI implementation; trust establishment, private key protection, CRL availability, key generation, legislation compliance. Table 3 describes how these elements were addressed in the UAE project.

Table 3. Key security issues in PKI addressed in the UAE project.

| Key Risks | Description |
|---|---|
| Trust establishment | The procedures followed to verify the individuals identities, before issuing identity certificates. The issuance of certificates is linked with the ID card enrolment process. Individuals go through vivid registration process that includes: biographical data capture, portrait, fingerprint biometrics capture, verification with civil and forensic biometric databases, biographical data verification with the Ministry of Interior's database and other black-listed lists. The certification revocation procedures are linked mainly with Ministry of Interior's database and strict policies and procedures. |
| Private key protection | The infrastructure is hosted in a highly secure physical location, that is ISO 27001 certified. |
| CRL availability | A list of serial numbers of all the digital certificates that have been cancelled; CRL (Certification Revocation List) to allow other institutes verify the status of any presented digital certificate, is designed for 24/7 availability and to maintain a strong and secure architecture to avoid security breaches and a comprehensive fail-over plan that provides a secondary in infrastructure to maintain availability of services in the case of failure of the primary infrastructure. |
| Key generation | Public and private keys of the certifying authority are generated using proprietary cryptographic algorithms. The user certificates are generated using market standard |

| | cryptographic algorithms. The technical key lengths are 2048, where as the user keys are 4096. |
|---|---|
| Legislation compliance | The government is currently working on developing the legal framework to recognise the operation of the PKI and the usage of digital certificates and digital signatures. International guidelines concerning PKI are being consulted such as (EU Electronic Signatures Directive, EU Data Protection Directive).<br><br>The UAE government issued a low on electronic transactions, however here is not legal act concerning the usage of digital certificates and signatures. |

## 5.2. Management Involvement: Shifting the Focus from PKI as a Technology to a Business Enabler

The adoption of PKI has the potential to deliver significant benefits to many sectors including e-Government, healthcare and banking. However, for such adoption to happen, it was important to understand and appreciate the business value, business requirements and business integration issues [37] relevant to potential PKI customers in the above mentioned example sectors.

As part of our strategy to implement a nation-wide PKI, it was seen important to consult potential customers across many sectors including e-government authorities both on the federal and local levels. In doing so, potential customers would see the benefits of PKI as a business enabler. We paid much attention to collect the necessary business requirements that would help tightening the future PKI functional requirements. See also Appendix-C.

Undoubtedly, deployment of a functioning PKI is extremely difficult in practice [7,8,36-38]. Weak understanding of the PKI technology by top management and lack of qualified resources in the field will always be a challenging factor. Before we reached to a consensus on the PKI design and functions, there was much confusion about the full scope of this project.

We noted that practitioners in the field of government identity management systems who are interested in PKI applications have deep-seated narrowed focus when thinking about such technologies. They tend to limit their focus purely on PKI services such as digital certification and electronic signatures in the context of e-government and e-commerce, without much comprehension of how PKI could be integrated with their business needs and practices. The aggressive marketing promises by private sector consultants and vendors have contributed somehow to some misconceptions in the minds of government officials of PKI applications.

Management involvement was important in some of the regular review meetings that required restating project objectives in a user friendly terms. It was common for the technical teams in the project to fell victims of technical-driven discussions and away from the global business objectives. It was important to remind the teams to reflect the interests of stakeholders in the government rather than just the interest of the implementing organization.

From a management standpoint, we tried to stop attempts of innovation as people tend to act sometimes in complex projects, and keep them focused on the business requirements, and the overall PKI functions. Stakeholders on the other hand needed several awareness sessions of the scope and deliverables of the project. It was important to visualize and present cases of how their applications will be integrated with PKI, and highlighting the immediate benefits. High attention was given to the development of Government-2-Citizen PKI enabled applications.

### 5.3. Implementation Approach

An agile but incremental phased implementation approach was followed, that emphasized the delivery of functionalities that could meet the immediate demands of local e-government authorities. The earlier workshops concentrated on discussing and refining business and technical requirements with the relevant stakeholders.

Specific attention was given to the development of the interfaces required to integrate the ID card system with the PKI solution, and allowing at the same time, agreement among the stakeholders on business and technical requirements. This allowed e-government authorities to experiment the authentication capabilities offered by the ID card including the online validation process.

This allowed the different groups in the organization to concentrate on the other building blocks of the PKI project, as they were running in parallel; such as technical workshops related to integration needs, testing, documentation, enforcement of policies, guidelines and compliance, digital signature laws, etc.

### 5.4. PKI Workflow and Lifecycle

It was important that we go through the full lifecycles of digital certificate-based identities, and how encryption, digital signature and certificate authentication capabilities are mapped to business needs and translated into real applications. See also [37,38]. It was also important to set clear procedures to handle smart card life cycle management requirements; renewal, replacement, revocation, unlocking, and the overall helpdesk and user support requirements.

Another issue that was considered in the PKI workflow was related to the incremental size of the Certificate Revocation List (CRL) which must be maintained and updated for proper validation of each transaction which occurs using the ID card certificate. An unverified transaction can provide important information or access for a potential intrusion. Thus, CRL is a significant security flaw in the operation of the PKI, and the maintenance of this list is one of the most strenuous challenges facing any CA.

The list of revoked certificates was envisaged to be well over multi gigabytes of size, and searching the list for invalid certificates will result in long delays as it will force some applications to forego a comprehensive check before carrying out a given transaction. Therefore, a Positive Certification List (PCL) was also implemented to avoid this challenge in the future.

### 5.5. Legal Framework

It was important that the PKI deployment is associated with a legal framework to regulate the electronic authentication environment and support the provision of online services in the public sector. See also [39]. The following items were key preparation issues addressed through intergovernmental working groups:

1. **Well-documented Certificate Policies (CPs) and Certification Practice Statements (CPSs)**. CPs and CPSs are tools that help establish trust relationships between the PKI provider, the subscribers (end-users) and relying parties (i.e. implementers of PKI-enabled applications).

2. **PKI assessment and accreditation** were seen as an important trust anchor, as it will determine compliance to defined criteria of trustworthiness and quality. Such assessment and audit was set as a prerequisite to be included in the trusted root CAs program. As part of the implementation of the future PKI applications, it was recommended that an assessment of the existing Certificate Policies (CPs) is conducted. This resulted in RFC 2527 compliant CPs and CPSs [40].

3. A **digital signature law** that would define the meaning of an e-signature in the legal context. The law needed to recognize a digital signature in signed electronic contracts and documents as legally binding as a paper-based contracts.

## 6. CONCLUSIONS

Public Key Infrastructure has proven itself invaluable in e-government and e-commerce environments despite the complexity and associated risks that may stem from its application. We observe that many of the current PKI projects have limited applications in e-government domain because it is mainly sponsored and managed by private sector organizations. Telecom companies in many countries in the Middle East region for example have implemented PKI systems but face challenges to expanding their limited user community.

Establishing and using a government based certification authority, would logically acquire higher levels of trust in the certificate issuance process and in the identities of the recipients of the certificates. The integration of PKI into central government identity management systems is believed to support the diffusion and acceleration of e-government progress, that is, the provision of citizen services and outreach over digital networks. The presented case study of the UAE PKI project and the approach the government has followed to integrate it part of its federal identity management system, was aimed to share knowledge and improve understanding of government practices in the field.

Assessment of the success of this proposed structure was beyond the scope of this article, as the implementation was undergoing during the preparation of this article. However, it will be published in a separate article once the full implementation is complete.

Without a doubt, the maturity of e-government requires significant efforts by both practitioners and researchers to support the development of horizontal and vertical e-government integration [41-43]. Governments need to prepare themselves to introduce social changes of work roles, attitudes and new competence needs. Governments are seen to be the entity responsible to lay down and develop the foundation of digital identities.

PKI remains a crucial component to provide higher security levels in digital forms, and will have a triple effect if integrated with the existing government trusted identity management systems. As the adoption of PKI in government projects is likely to continue, opportunities exist for future researchers to examine the success of such implementations.

**About the Author**

Dr. Ali M. Al-Khouri is the Director General of Emirates Identity Authority in the United Arab Emirates. He has been working in the government sector for more than 20 years. He has graduated from top leading UK universities. He received his B.Sc. (Hons.) in Business IT management from Manchester University; M.Sc. in Information Management from Lancaster University; and EngD in Strategic and Large Scale Projects Management in Public Sector from Warwick University. His research interest areas focus on developing best practices in public sector management and the development of information societies.

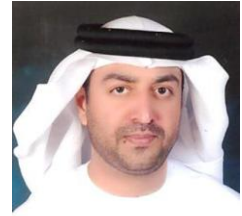

# APPENDIX-A: MAJOR PKI PROJECTS WORLDWIDE

1. **ICAO PKD (International Civil Aviation Organization Public Key Directory)**

   This is a global PKI directory implementation for achieving interoperable ePassports worldwide. The key benefit of this project is the PKI Validation of ePassport.

   This project allows border control authorities to confirm that:

   - The ePassport document held by the traveler was issued by a bonafide authority.
   - The biographical and biometric information endorsed in the document at issuance has not subsequently been altered.
   - Provided active authentication and / or chip authentication is supported by the ePassport, the electronic information in the document is not a copy (ie clone).
   - If the document has been reported lost or has been cancelled, the validation check can help confirm whether the document remains in the hands of the person to whom it was issued.

2. **SWIFT – PKI at application level (SWIFTNet PKI), and another PKI at network level (VPN)**

   SWIFT's public key infrastructure (SWIFTNet PKI) service issues digital certificates to financial institutions and corporates, thereby enabling a trusted, provable and confidential end-to-end communication over SWIFTNet.

   In addition SWIFT's VPN PKI issues certificates to its network infrastructure to secure all network traffic using VPN protocols.

3. **European TLIST of the 27 Member States**

   On 16 October 2009 the European Commission adopted a Decision setting out measures facilitating the use of procedures by electronic means through the 'points of single contact' under the Services Directive. One of the measures adopted by the Decision consisted in the obligation for Member States to establish and publish by 28. 12.2009 their Trusted List of supervised/accredited certification service providers issuing qualified certificates to the public. The objective of this obligation is to enhance cross-border use of electronic signatures by increasing trust in electronic signatures originating from other Member States. The Decision was updated several times since 16.10.2009, the last amendment was made on 28.7.2010

   The EU Trusted Lists benefits above all to the verification of advanced e-signatures supported by qualified certificates in the meaning of the e-signature directive (1999/93/EC) as far as they have to include at least certification service providers issuing qualified certificates. Member States can however include in their Trusted Lists also other certification service providers.

Member States had to establish and publish their Trusted List by 28.12.2009 at least in a "human readable" form but were free to produce also a "machine processable" form which allowed for automated information retrieval. In order to allow access to the trusted lists of all Member States in an easy manner, the European Commission has published a central list with links to national "trusted lists"

(https://ec.europa.eu/information_society/policy/esignature/trusted-list/tl-hr.pdf).

4. **ERCA (European Root Certification Authority)**

The main ERCA deliverables are the Member State Authority [MSA] policy review and the Key generation for Member State Authority CAs. During an ERCA signing session, countries receive the symmetric and asymmetric encryption keys for use by their Member State Authority.

The Member State Authorities will issue certificates on smart cards required for the operations of the tachograph which a device that records a vehicle's speed over time, monitor driver's working hours and ensure that appropriate breaks are taken.

5. **TSCP PKI (Transatlantic Secure Collaboration Program)**

TSCP program involves leading aerospace & defense companies in the USA and Europe including Boeing, BAE Systems, EADS, Lockheed-Martin, Northrop Grumman, Raytheon and Rolls-Royce. Supporting Governments include the US DoD, the UK MoD and the Government of Canada.

The challenge addressed by TSCP PKI is the increasing reliance on the electronic creation, transmission and manipulation of information in order to meet schedule and efficiency objectives. The emerging business environment requires that this occur with an international workforce subject to multiple jurisdictions. This presents significant business risk to the companies involved, in terms of compliance with national laws and regulations on data transfer, increased complexity of governance and oversight, and IT security.

The TSCP PKI represents the bridge CA that allows interoperability between the program members CAs.

6. **FBCA - US Federal bridge CA**

The main objective of this bridge CA is to provide one CA for cross-certification between main CAs in the US and avoid complicated mesh cross-certification. FBCA is designed to create trust paths among individual PKIs. It employs a distributed and not a hierarchical model.

The FPKISC, Federal Public Key Infrastructure Steering Committee, oversees FBCA development and operations including documentation, enhancements and client-side software. The FBCA operates in accordance with FPKI Policy Authority and FPKISC directions.

FBCA is in charge of propagating policy information to certificate users and maintain a PKI directory online 24 X 7 X 365.

# APPENDIX-B: TOOLKIT CAPABILITIES

| Toolkit Function | Require Secure Messaging? | Require Online Validation Service from EIDA? | What kind of validation is required? |
|---|---|---|---|
| **Read Public Data** | No | No | Public data files (read from the card) are signed. The Toolkit function "**Read Public Data**" verifies the signature on these file as part of the reading process. The verification process happens locally on the end user environment and as such it is an offline process that does not require any online additional service from EIDA. |
| **Authentication (with PKI)** | Yes<br><br>The justification is as follows:<br>• Firstly, the authentication process involves the PKI applet where the authentication key pair (and corresponding certificate) is stored<br>• The ID card (actually the PKI applet) requires PIN verification prior to authorizing the usage of the authentication key pair.<br>• PIN verification requires secure messaging with the PKI applet. | PKI authentication requires the following steps:<br><br>- Secure Messaging with the PKI applet<br>- PIN Verification<br>- Authentication process through which the cardholder authenticate certificate is validated.<br><br>Establishing a Secure Messaging with the PKI applet of the ID card does not require being online with EIDA. This is due to the fact that the SAGEM PKI DLL part of the Toolkit contains the keys that enable establishing a Secure Messaging with the PKI applet locally (**offline process**).<br><br>Regarding certificate validation, EMIRATES ID does as part of the PKI solution an online service (OCSP) that offers real time verification of certificate status. Whether the Service Provider business application will use this service depends basically on their will to rely all the time on an **online service** for certification validation. The alternative for them would be to download CRLs | Authentication certificate validation is a pre-requisite to complete the authentication process. Two modes are available for certification validation:<br><br>1. Using CRLs: in this case, CRLs are downloaded by the business application regularly. However the actual processing of certificate validation is an offline process that happens locally on the end-user's environment.<br><br>2. Using EMIRATES ID OCSP server: This is an online service that relieves the business application from the complex processing of CRLs and provides real time validation of certificate revocation status. |

| | | frequently from the CA repository and to process certificate and CRLs locally (**offline process**). | |
|---|---|---|---|
| **Biometric Match-Off-Card** | **Yes**<br><br>The Match-Off-Card requires reading the fingerprint from the ID card which is protected data that requires Secure Messaging with the ID Applet. | It depends on the business application architecture and deployment channel. The possible options are as follows:<br><br>1. The business application is offered as an online service (e.g. ADSIC e-services portal, e-services kiosks). In this case, the business application relies on an **online service** from EMIRATES ID that enables setting a Secure Messaging session with the ID applet.<br><br>2. The business application is deployed on user sites (e.g. municipalities) that are visited by end-users. The application deployed on user sites has a dedicated SAM device connected to it. Therefore Secure Messaging with the ID applet can be established using the SAM. The process is **offline** and does not require online connectivity to EIDA.<br><br>3. The business application is deployed on alternative channel such as kiosks. The kiosk would have an integrated SAM device and the processing is therefore **offline** and is similar to the 2$^{nd}$ point above.<br><br>4. The business application is deployed on standalone/**offline** devices such as Handhelds with integrated SAM. | No validation is required as part of the Off-Card-Biometric apart from performing the actual verification process that involves the fingerprint template read from the ID card. The verification happens locally on the end-user environment and does not require online connectivity to EIDA. |
| **Biometric Match-On-Card** | **Yes.**<br><br>The Match-On-Card requires an interaction with the MOC applet of the ID card. The overall process for Match-On-Card can be summarized as follows:<br><br>1. Setup a Secure Messaging Session with the MOC | By definition, the Match-On-card process is an **offline process** that shall not require online connectivity to EIDA. Therefore, the MOC process is typically used in situations where the business application has a connected SAM. Examples of such a deployment would be:<br><br>1. The business application is deployed on user sites (e.g. municipalities) that are visited by end-users. The application deployed on user sites has a dedicated SAM device connected to it. Therefore Secure Messaging with the MOC applet can be established using the SAM. | No validation is required as part of the On-Card-Biometric apart from performing the actual verification process that happens locally on the end-user environment and does not require online connectivity to EIDA. |

| | | | | |
|---|---|---|---|---|
| | Applet of the ID card<br><br>2. Perform the actual MOC verification after capturing the end-user fingerprint | 2. The business application is deployed on standalone devices such as Handhelds with integrated SAM. | | |
| **Digital (Transaction) Signature** | **Yes.**<br><br>Justification is similar to the Authentication process with PKI. | The digital signature verification process by the business application involves the following steps:<br><br>1. Signature Verification<br>2. Certificate Path Build<br>3. Certificate Path validation (where certification revocation status is checked)<br><br>EMIRATES ID offers 2 online services to business partners.<br><br>- Online certificate validation through the OCSP server. The discussion on this is similar to the one provided for the Authentication process with PKI (see Authentication entry).<br><br>- Online Signature Validation: In this case the whole signature verification process is outsourced to EIDA. | Authentication certificate validation is a pre-requisite to complete the authentication process. The two modes discussed under Authentication with PKI are application (see Authentication entry above).<br><br>If the business partner decides to use EMIRATES ID **online service** for signature validation, then this requires an online connectivity to EIDA. | |

## APPENDIX-C: SUMMARY OF BUSINESS REQUIREMENTS AND PKI FUNCTIONAL REQUIREMENTS

| | Concept | Description | Opportunities | Challenges |
|---|---|---|---|---|
| **EMIRATES ID as the PKI provider for the UAE e-Government** | Only CAs operated by EMIRATES ID are recognized by stakeholders in the eGovernment sector. | EMIRATES ID is already managing the PKI for the eID card issuing project (i.e. population CA). This PKI will be upgraded including the implementation of validation services (CRL, OCSP). Future UAE eGovernment PKI projects will take advantage of the new PKI to be implemented by EIDA. Finally, existing EMIRATES ID PKI applications (including the eID card issuing project) will be migrated to the future EMIRATES ID PKI. | • EMIRATES ID being the main PKI provider in the UAE and the recognized one for eGovernment projects<br>• EMIRATES ID becomes a revenue driven organization | • Lack of support from Stakeholders |

| | | | | Pros | | Cons |
|---|---|---|---|---|---|---|
| **Offering Managed PKI services** | EMIRATES ID to cross-certify other government and commercial CAs and to offer managed PKI services for these. | EMIRATES ID will offer Managed PKI services aimed at enterprises planning to establish a Certification Authority (CA) for addressing their particular business needs. Organizations that might be interested in such services are banks and governmental organizations (e.g. healthcare, education). Also EMIRATES ID can cross-certify (acting as a root) other CAs. | | • Establish EMIRATES ID credibility as the trusted PKI provider in the UAE<br>• An additional revenue stream for EIDA | | • More operational and infrastructural requirements<br>• EMIRATES ID diversifying into a business stream unrelated to EMIRATES ID core business |
| **Supporting encryption certificates** | EMIRATES ID PKI to support issuing encryption certificates with key backup | EMIRATES ID can establish CA for encryption certificate issuing with key escrow. Each encryption key pair will be issued under the control of the certificate holder of the organization to which he belongs. | | • Full range of certificate types offered to EMIRATES ID potential customers<br>• Support the eID card issuing project in case encryption certificates are needed | | • Maintaining and backing up keys is a liability issue and additional operation overhead<br>• Encryption tend to be seen as a threat to national security |
| **Supporting multi purpose certificates** | PKI to support the issuance of multi purpose certificates | EMIRATES ID PKI will support issuing and managing different types of certificates such as certificates for:<br>• VPN devices<br>• Web servers (SSL certificates)<br>• Simple Certificate Enrolment Protocol (SCEP) devices<br>• Attribute (role-based) certificates<br>• Certificates for professionals (e.g. doctors)<br>• Anonymity certificates (e.g. by omitting first and last names from the certificate)<br>EMIRATES ID PKI will offer the enrolment methods needed to issues these types of certificates by the relevant organizations. | | • Support PKI-enabling in the UAE by providing certificates that fulfill the requirements of different sectors<br>• Full range of certificate types offered to EMIRATES ID potential customers<br>• Additional revenue streams | | • Eventually additional CAs and CPs to manage |
| **Promote electronic signatures law** | promote electronic signatures law | EMIRATES ID will define an e-Signature law that will define the legal framework for electronic signature as well as set EMIRATES ID as the regulatory authority (root CA) accrediting and certifying other organizations. | | • Promote digital signature and eID card usage in the UAE<br>• Establish EMIRATES ID as the regulatory authority | | • Citizen reaction to be legally liable for digitally signing documents |

| | | | | responsible for the secure PKI usage in the UAE | |
|---|---|---|---|---|---|
| **Trusted Time-stamping** | provide trusted time-stamping service (RFC 3161-compliant)<br><br>(RFC 3161 is Time-stamping protocol) | The time-stamping services will allow EMIRATES ID enforcing and offering non-repudiation services so that a signature remains valid long-term after it has been created. Time-stamping general added-value is to provide an irrefutable proof that a document existed at a certain point in time. | • Enforcing non-repudiation services (long-term validity of e-documents)<br>• Additional revenue stream<br>• Establish EMIRATES ID position as the trusted PKI provider in the UAE | • Time-stamping servers rely on the availability of a trusted time source (e.g. NTP server)<br>• Time-stamping is an online service, therefore EMIRATES ID need to adhere to SLAs defined for its customers |
| **Online Validation Services** | to support Online Certificate Status Protocol (OCSP) | OCSP allows providing timely secure information on certificate revocation status. | • Timely access to certificate revocation status information<br>• Easier PKI-enabling compared to relying on CRLs only<br>• Eventually additional revenue streams<br>• Simpler integration models with e-Government projects<br>• Establish EMIRATES ID position as the trusted PKI provider in the UAE | • OCSP is an online service, therefore EMIRATES ID need to adhere to SLAs defined for its customers<br>• |
| **Long-term Archive Services (LTA)** | to provide/promote long-term archive services (LTA) | EMIRATES ID will provide/promote long-term archive services (LTA) which will enable the preservation of data integrity over the time. The LTA service will be particularly useful with signed documents whose validity shall be preserved over time. | • Enforcing long-term non-repudiation services offered by EIDA<br>• Additional revenue stream<br>• Establish EMIRATES ID position as the trusted PKI provider in the UAE | • additional operation costs for EIDA<br>• relatively complex IS<br>• Operating the LTA might be out of scope for EMIRATES ID as e-Identity authority |
| **E-Notary services** | E-Notary services promoted/provided by EIDA | An e-notary is a PKI based application that allows adding trust in digital transactions (i.e. such as e-commerce transactions). Such services provides | • Facilitating and enabling trusted e-commerce transactions | • Typically e-Notaries applications are bespoke<br>• No successful |
| | | guarantee to the parties involved in the transaction that they can trust each other and provides the proofs needed to establish that a transaction took place. Potential customers for such services are e-Justice and e-Commerce sectors. | • Additional revenue stream for EIDA | implementation (e.g. lessons learned) worldwide so far<br>• Current e-commerce law does not define what an e-Notary stands for<br>• Lack of adoption of e-commerce implies the lack of adoption of an e-Notary application |
| **eID starter kit for citizens (certificate holders)** | will provide an eID starter kit | To support using of PKI, EMIRATES ID will create and provide a package that will facilitate the installation of everything needed for using the eID card by the cardholder. | • Simple adoption of PKI and eID card by citizens and residents<br>• Promote the usage of the eID card within the UAE<br>• Strengthen the position of EMIRATES ID as the promoter of the eID card in the UAE<br>• Promoting the upgrade of the existing PCs park in the UAE | • Lack of IT knowledge among citizens<br>• Existing PCs park not supporting the installation of the package to be installed by the citizen<br>• Expensive operation in case EMIRATES ID targets additional platforms (e.g. UNIX) other than Microsoft Windows<br>• Flooding of the EMIRATES ID helpdesk when the roll-out of this package starts |
| **eID development kit** | EMIRATES ID will provide an eID development kit for eID (e.g. PKI) applications developers | As opposed to the eID starter kit used by citizens, the eID development kit will be used by whoever will be interested in developing eID and PKI enabled applications. The kit includes code samples, APIs, access to EMIRATES ID test PKI infrastructure, sample eID cards, etc. | • Quick adoption of PKI by relying parties (i.e. by the implementers of<br>• Trigger for EMIRATES ID to implement a fully-fledged PKI in development environment to service PKI applications implementers | • The toolkit is useless if EMIRATES ID does not target the right platforms<br>• Lack of EMIRATES ID experienced staff to support the organizations using the kit |
| | | | • Additional revenue stream for EIDA | |
| **Third party PKI modules/application type approval (i.e. certification)** | Third party PKI and eID application certification | EMIRATES ID will help the developers of eID applications promoting their applications. This includes testing and the type approval of their applications so that they get accredited as authorized to process eID cards. | • Enable the implementation of secure eID-enabled applications<br>• Build confidence of citizens (eID card holders) in business applications using the ID card<br>• Establish EMIRATES ID position as the trusted party around eID-enabling | • Testing applications requires lots of skills (including PKI, eID and testing skills). People having these skills shall be available<br>• Liability issue for EMIRATES ID in case security holes are found in some applications (that did receive type approval from EIDA) |
| **Certified mailbox service** | EMIRATES ID to implement and operate a certified mailbox service | Such service would allow each citizen/resident to have a certified mailbox (of the form firstname.lastname@mymailbox.ae) where we would receive official courier from governmental organization. The mailbox could also be used to receive any useful courier (including bills, etc). | • Business case application for using the eID card<br>• Promoting G2C and B2C markets<br>• Additional income streams<br>• Government being closer to citizens<br>• New business stream for commercial organizations | • Citizens may find that such service is a threat against their privacy<br>• Liability issue for EMIRATES ID in case a threat agent breaks into one mailbox<br>• Lack of adoption by citizens unless the service is mandatory |
| **Card Validation Gateway** | Implement an access control server for eID cards | The Validation Gateway maintains a "hotlist" of cards that has been temporarily or permanently blocked. Such service could be useful for specific public services such as police and kiosks. The Validation Gateway is also needed in case EMIRATES ID implements Post Issuance Personalization (PIP) of card applications. | • Support post-issuance personalization of eID cards<br>• Providing a trusted gateway responsible for providing a blacklist of revoked cards<br>• Easier integration models between EMIRATES ID and eID applications implementers | • The Validation Gateway can be seen as a single point of attacks by threat agents<br>• The validation gateway being developed without thinking the right integration interfaces with eID enabled applications |